# Maximum Matchings in Graphs for Allocating Kidney Paired Donation

### Sommer Gentry
Mathematics Department, United States Naval Academy, Annapolis MD 21402, gentry@usna.edu

### Michal Mankowski
Computer, Electrical and Mathematical Sciences and Engineering Division , King Abdullah University of Science and Technology, Thuwal, Saudi Arabia, michal.mankowski@kaust.edu.sa

### T. S. Michael
Mathematics Department, United States Naval Academy, Annapolis MD 21402, tsm@usna.edu

### Dorry Segev
Department of Surgery, Johns Hopkins University School of Medicine, Baltimore MD 21287, dorry@jhmi.edu

Relatives and friends of an end-stage renal disease patient who offer to donate a kidney are often found to be incompatible with their intended recipients. Kidney paired donation matches one patient and his incompatible donor with another patient and donor in the same situation for an organ exchange. Let patient-donor pairs be the vertices of an undirected graph $G$, with an edge connecting any two reciprocally compatible vertices. A matching in $G$ is a feasible set of paired donations. We describe various optimization problems on kidney paired donation graphs $G$ and the merits of each in clinical transplantation. Because some matches are geographically undesirable, and the expected lifespan of a transplanted kidney depends on the immunologic concordance of donor and recipient, we weight the edges of $G$ and seek a maximum edge-weight matching. Unfortunately, such matchings might not have the maximum cardinality; there is a risk of an unpredictable trade-off between the quality and quantity of paired donations. We propose an edge-weighting of $G$ which guarantees that every matching with maximum weight also has maximum cardinality, and also maximizes the number of transplants for an exceptional subset of recipients, while reducing travel and favoring immunologic concordance.

## 1. Introduction

The preferred treatment for end-stage renal disease is kidney transplantation, but there are not enough donor kidneys available to meet the overwhelming need. As of July, 2007 there are 82,752 candidates in the United States waiting for a kidney (UNOS 2009). Often a family member or a friend offers to donate one of his two kidneys, but approximately one-third of such offers are ruled out because the donor's blood or tissue types are incompatible with the intended recipient. Kidney paired donation circumvents these barriers by matching an incompatible pair to another pair with a complementary incompatibility (Rapaport 1986, Montgomery et al. 2005). In simultaneous operations, the donor of the first pair gives to the recipient of the second pair, and vice versa. Each donor's surgery occurs in the same hospital as the actual recipient, so in a match between distant hospitals, at least two people must travel.

Recently in the operations research and economics literature, investigators have used optimization models for both the transplantation decisions of individuals [Alagoz et al. (2004) for liver transplants, Su and Zenios (2005) for kidney transplants] and the societal allocation of organs [Zenios (2002) and Roth et al. (2005b) for kidneys, Stahl et al. (2005) for livers]. This paper takes the perspective of a central planner in determining a societally optimal allocation for kidney paired donation.

In the U.S., the United Network for Organ Sharing (UNOS) is charged with allocating organs from deceased donors, which it does according to a points system that ranks recipients on the waiting list for kidneys. On the other hand, live donation has for the most part been directed donation, in which the donor is willing to give only to his specified recipient. Kidney paired donations must be arranged by some clearinghouse to satisfy the reciprocal compatibility constraint: a donor will give only if his intended recipient will receive a kidney. A kidney paired donation allocation assigns every incompatible pair either to an exchange opportunity with another incompatible pair or to the group of unmatched pairs. Optimal kidney paired donation can be formulated as a maximum matching problem in a weighted graph for two-way paired donation, or more generally as an integer program if more than two pairs may be involved in any exchange.

This paper focuses on two-way paired donation, although paired donations involving three incompatible pairs have been accomplished at several transplant centers, and more transplants might be possible if three-way paired donations were permitted (Saidman et al. 2006, Roth et al. 2005a). In the medical setting, three-way paired donations are logistically complex and are more likely than two-way paired donations to be scuttled by the discovery of an unpredicted incompatibility or by a medical event affecting one of the donors or recipients. In its proposal for a United States paired donation registry, UNOS initially restricts the system to two-way paired donations based on these clinical considerations (UNOS 2006b). Optimizing $k$-way paired donations for $k > 2$ requires an integer programming formulation, which we discuss in Section 7.3.

## 1.1. Purpose and Outline

The contributions of the paper are: an overview of research in kidney paired donation allocation; the design of clinically suitable optimization models for paired donation; and some technical results. A paired donation allocation corresponds to a matching in a graph which may have positive integer weights on either its vertices or edges. We distinguish between maximizing the sum of vertex weights and maximizing the sum of edge weights, and argue that edge weights are necessary to capture important features of clinical paired donation such as travel distance and risk of immunologic incompatibility. One desirable property of a maximum vertex-weight matching is that it simultaneously maximizes the number of transplants performed. Maximizing the sum of edge weights, on the other hand, does not necessarily maximize the number of transplants performed. That is, accounting for important factors such as the compatibility of each recipient with his donor may reduce the total number of transplants. For maximum edge-weight matchings, we give a constant-factor lower bound on the suboptimality of the number of transplants performed. We also give a constraint on the edge weights that guarantees that a maximum edge-weight matching simultaneously maximizes the number of transplants performed. We extend this result to consider a subgroup of recipients for whom a transplant is medically urgent, and give an edge weighting that guarantees that a maximum edge-weight matching simultaneously maximizes both the number of medically urgent recipients transplanted and the total number of transplants performed. Our proofs rely on elementary inequalities for matchings in edge-weighted graphs.

Although the concept of kidney paired donation was described in 1986 (Rapaport 1986), only about 700 U.S. patients have received paired donations, all since 2001. Creation of a national kidney paired donation registry in the U.S. has been delayed more than 20 years, partly because stakeholders did not believe that logistical problems like donor travel requirements could be overcome. We present a method for allocating matches in a national kidney paired donation registry that simultaneously (a) achieves the absolute maximum number of paired donations; and (b) reduces the number of long-distance exchanges and favors better immunologic (human leukocyte antigen, or HLA) concordance. Further, we present a second allocation method to handle high-priority recipients that (a) achieves the absolute maximum number of donations for the exceptional recipients; (b) achieves the absolute maximum number of paired donations; and (c) reduces the number of long-distance exchanges and favors better immunologic concordance for each recipient. These methods convert preemptive multi-objective optimal matching problems into maximum edge-weight matching problems.

In addition to providing the two methods described, we hope to introduce the reader to the breadth of theoretically and practically interesting problems in kidney paired donation. We have two audiences in mind: the operations research community, who may verify both the utility of graph models for allocating kidney paired donations, and our proofs that the maximum edge-weight matchings possess the desired optimality properties; and transplant professionals, who may find the allocation methods of great practical significance.

The remainder of the paper is organized as follows: Section 2 introduces some clinical and practical aspects of paired donation. In Section 3 we consider various formulations of the objective for allocating kidney paired donation using maximum matchings. In Sections 4 and 5, we establish our main results, which give useful guarantees about the number of edges in a maximum edge-weight matching when suitable restrictions are placed on the edge weights of a graph. Our suggested method for allocating kidney paired donation is detailed in Section 5.2. Computational results are presented in Section 6. Section 7 describes graph models for more general forms of kidney donor exchange, especially the case of $k$-way donor exchange where $k > 2$. We discuss limitations of our analysis in Section 8.

## 2. Clinical Aspects of Paired Donation

This section introduces the medical and logistical complexities of kidney transplantation in general and paired donation in particular.

### 2.1. Determining Reciprocal Compatibility Between Pairs

Medically, compatibility is determined both by blood type (O, A, B, or AB) and by existing tissue antibodies. A blood type O recipient can accept only a blood type O donor; a blood type A recipient can accept only a blood type O or A donor; a blood type B recipient can accept only a blood type O or B donor; and a blood type AB recipient can accept a donor of any blood type. Even if a donor has a compatible blood type, the recipient may be *sensitized* to the donor, meaning that the recipient has preformed tissue antibodies that will attack the donor's kidney. A laboratory crossmatch test, performed before any kidney transplant, will be positive if the recipient is sensitized to his donor. A woman can become sensitized to her husband and children via pregnancy. A highly sensitized recipient is one who has preformed antibodies to most of the population and so has difficulty finding any compatible donors.

At a very few centers, programs exist to desensitize recipients so that they can receive kidneys from blood type incompatible or tissue incompatible donors. Desensitizing recipients involves greater risk and cost than compatible transplantation, and the difficulty of desensitizing varies with the particular donor and recipient. Desensitization may be used in conjunction with paired donation, so that someone with a stubborn incompatibility to his own donor is matched to an incompatible but more easily-desensitized-for donor (Montgomery et al. 2005).

Recipient, donor, and transplant center restrictions must also be considered in determining whether two pairs can exchange donors in a paired donation. Recipients might refuse certain donors because of donor age or medical conditions. Donors might refuse certain matches because of the distance they would need to travel for donation, or because the center that would perform their donation does not offer the less invasive laparoscopic donor operation. Transplant centers differ in their willingness to operate on hypertensive or older donors. Registries should document these restrictions to maximize the chances that the eventual allocation will be accepted.

## 2.2. Simulations of Recipients and Incompatible Donors

U.S. transplant centers must report medical and demographic information concerning both potential organ transplant recipients and actual organ donors to UNOS through the Organ Procurement and Transplantation Network (UNOS 2006a). However, there is no requirement and no mechanism for reporting about willing live kidney donors who are incompatible with their intended recipients. Indeed, because blood type compatibility problems are widely known, but paired donation is not, many willing donors may be "ruled out at the dinner table". Researchers in kidney paired donation rely instead on simulated databases of patients with their incompatible donors (Segev et al. 2005a, Zenios et al. 2001, Saidman et al. 2006). Most living donors are genetically related to their recipients, so our studies include simulated inheritance of both blood type and HLA antigens (Gentry et al. 2005). We compared our simulated patient-donor blood type distributions to registration data for the Netherlands kidney paired donation registry (de Klerk et al. 2005) using a Pearson's chi-square test, and found no evidence that our simulated blood types differ from the observed blood types (p=0.65). Although we refer to patient-donor pairs, it is best that each recipient bring forward as many willing donors as are available. Only one of a recipient's volunteers will donate if a match is found.

## 2.3. Accumulation of Pairs

In practice, recipients and their incompatible donors present to physicians on an ongoing basis. Zenios (2002) described an optimal controller for dynamic assignment of incompatible pairs to either kidney paired donation or list paired donation (see Section 7.2). However, if every feasible paired donation were performed immediately, there would be no opportunity to take advantage of optimal matching algorithms, and fewer transplants overall could be performed. In this paper, we view the dynamic problem as a static optimization problem. The transplant community has recognized the need for a waiting period of three to six months during which incompatible pairs accumulate in advance of a *match run* that would solve a static optimization problem.

The effectiveness of kidney paired donation programs depends on the participation of a sufficient number of incompatible pairs. For instance, we project that the percentage of people who match for paired donations is about 22% in pools of 15 pairs, and 38% in pools of 250 pairs (Gentry et al. 2007). More than half of the participants will not match, primarily because of a blood type imbalance among incompatible pairs. Blood type O recipients can only accept a blood type O kidney, but blood type O donors can give to recipients of any blood type. Thus there will be a relative scarcity of blood type O donors among incompatible pairs.

## 2.4. Medical and Ethical Priorities in Allocating Transplants

It is tempting to believe that one could define an uncontroversial utility criterion, such as maximizing quality-adjusted-life-years, for the allocation of kidney paired donation matches. Although maximizing quality-adjusted-life-years is rational for an individual making medical decisions, in a centralized allocation system the years of life resulting from transplants would benefit some individuals and not others. Ensuring an equitable allocation might conflict with ensuring efficiency. For instance, it would not be acceptable to deny transplantation to a particular ethnic minority, even if survival after transplantation is lower for members of that minority than for other populations. As another example, giving priority to recipients who have themselves been prior live donors seems an equitable recompense for the gift they have given, regardless of whether such a priority increases efficiency by encouraging more people to become live donors.

The existing allocation system for deceased donor kidneys can be viewed as a compromise between efficiency and equity; see Zenios et al. (1999) for a detailed examination of the trade-off between these objectives in the allocation of deceased donor kidneys. The deceased donor system (UNOS 2006c) gives priority, for example, to candidates who are difficult to match (*sensitized*) or who are children, and gives infinite priority to any kidney allocation with perfect immunologic concordance (*zero-mismatch*). For an account of conflict between individual autonomy and system-wide utility in the deceased donor allocation system, see Su and Zenios (2005). We will say more about the choice of optimization objective for kidney paired donation in Section 3.

## 2.5. Efforts to Create Kidney Paired Donation Registries

The U.S. currently lacks a centralized kidney paired donation registry. A few regional consortia exist, and several transplant centers maintain lists of recipients with incompatible donors. The great promise of kidney paired donation led to the passage of a bill removing legal barriers to a national paired donation matching registry (U.S. Senate and U.S. House of Representatives 2007). Also, UNOS has released a detailed proposal for a kidney paired donation registry (UNOS 2006b). A U.S. national registry is projected to increase the rate of live donation substantially. A registry will also save the medical system hundreds of millions of dollars (Segev et al. 2005b). The government will realize financial benefits from paired donation because dialysis automatically qualifies patients for Medicare, and because dialysis is more costly than transplantation.

Both South Korea (Park et al. 1999) and the Netherlands (de Klerk et al. 2004) currently operate national kidney paired donation registries. The Korean registry models allocation as a preemptive multi-objective optimization problem and solves it by enumerating all feasible allocations (Kim et al. 2007). While enumeration is a correct solution method, it will fail due to combinatorial explosion if the registry grows to even a moderate size (the largest pool reported had 39 patient-donor pairs). The matching procedure used in the Netherlands does not guarantee an efficient matching; rather, it uses a greedy algorithm to add edges to the allocation in order of increasing match probability of both recipients, as defined in Keizer et al. (2005). One may construct an example in which this procedure matches a set $R$ of recipients even though it is possible to match all recipients in $R$ and some other recipients, too. However, we can not determine whether this has actually occurred. Canada's national kidney paired donation registry began operation in 2008, and uses an edge-weighted graph optimization model as elaborated in this manuscript.

## 3. Objectives for Optimal Matchings in KPD Graphs

We represent reciprocal compatibility between incompatible pairs by edges in an undirected graph $G$. Each vertex of $G$ represents an incompatible patient-donor pair, and each edge represents a feasible match. That is, there is an edge between two vertices of $G$ whenever the donor of the first pair is compatible with the recipient of the second pair, and also the donor of the second pair is compatible with the recipient of the first pair. We refer to $G$ as a *KPD graph*.

### 3.1. Matchings in Kidney Paired Donation Graphs

A *matching $M$* in a graph $G$ is a set of edges in $G$ such that every vertex of the graph is incident with at most one edge of $M$. The *matching number $\mu$* of $G$ is the maximum number of edges in a matching in $G$. There is a vast literature on matchings, matching numbers, and their applications. (See the classic paper by Edmonds (1965), the book by Lovász and Plummer (2009), and the survey in Pulleyblank (1995).)

Any feasible allocation of kidney paired donations within a KPD graph $G$ is a matching. A KPD graph is not bipartite in general, since any incompatible pair may in theory be reciprocally compatible with any other.

It is likely that individuals will have slight preferences among their feasible donors. A *stable matching* is one in which no two participants prefer each other to their matched partners. Suppose the registry matches $a$ with $b$ and $c$ with $d$, but $b$ prefers $c$ to $a$, and $c$ prefers $b$ to $d$. Then the registry has produced an unstable matching. Theoretically, $b$ and $c$ might leave their allocated matches and instead match with each other for a kidney paired donation. Although stable matchings are desirable in many applications of matching theory (Gusfield and Irving 1989), we claim that it is neither possible not particularly useful to require stability of the allocation matching for kidney paired donation. First, some graphs might not possess a stable matching. Gale and Shapley (1962) give an example. Second, the pairs are unlikely to have enough information to articulate their preferences, or to arrange a paired donation outside of a sanctioned registry. A laboratory and the complicity of the medical establishment are required even to determine whether two pairs are mutually compatible, to say nothing of the operations.

There are several notions of optimality for the matching $M$. One approach is to view all kidney paired donations as equally valuable. Then the matching $M$ is optimal provided it has the maximum cardinality $\mu$. However, maximizing the number of transplants is only one of many goals that physicians wish to achieve in allocating kidney paired donation. We now discuss two variant notions of optimality in which *weights* (positive real numbers) are assigned to the vertices or edges of $G$ to signify preferences among matchings.

The *vertex-weight* of a matching is the sum of the weights of the incident vertices. The *edge-weight* of a matching is the sum of the weights of its edges. If weights are assigned to the vertices of the KPD graph, then the objective will be to obtain the maximum vertex-weight matching. Otherwise, if weights are assigned to the edges, then the objective will be to obtain the maximum edge-weight matching. Edge-weighted matchings include vertex-weighted matchings as a special case when the weight for each edge is defined as the sum of the weights of the vertices connected by that edge. Furthermore, edge-weighted matchings have an advantage over vertex-weighted matchings in KPD graphs.

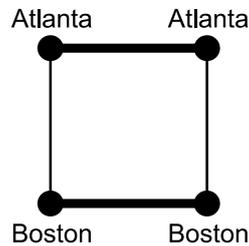

**Figure 1**    A kidney paired donation graph for four patient-donor pairs

Consider the simple KPD graph $G$ in Figure 1. Two distinct matchings in $G$ use all four vertices. A vertex-weighted scheme must fail to differentiate between the matching that includes two long-distance exchanges between Boston and Atlanta, and the matching that includes only local exchanges within the two cities, since all four vertices are included in either matching. An appropriate edge weighting of $G$, however, would assign larger weights to the bold edges so that the maximum edge-weight matching is the one that involves local exchanges within the two cities.

## 3.2.  Factors that can be captured using vertex weights

To express priorities among patients, Roth et al. (2005b) propose assigning a weight to each vertex of $G$. Organ allocation policy has long recognized several special categories of transplant candidates: pediatric candidates, the medically fragile, or candidates disadvantaged by restricted compatibility. A medically fragile candidate is one whose need for a transplant has become urgent. In the disadvantaged category, a highly sensitized candidate has a wide range of existing antibodies that make the search for a compatible donor like looking for a needle in the haystack. Physicians recognize at least two objectives: maximizing transplants for prioritized patients, and maximizing transplants overall. There is no trade-off necessary to maximize both, because maximum vertex-weight matchings always have maximum cardinality, as we show in Section 4.1.

### 3.3. Factors that can only be captured as edge weights

The desirability of a particular paired donation allocation actually depends on both edge properties and vertex properties. To express priorities among feasible kidney exchanges, we have proposed assigning a weight to each edge of the KPD graph $G$ (Segev et al. 2005b). Of course, a maximum edge-weight scheme can also consider factors related to vertices, such as the pediatric or prior live donor status of a recipient, by adding the weight attributable to a vertex property to the weight of every edge incident on that vertex.

We argued above that travel requirements are an edge property that cannot be captured using vertex weights. This section discusses this and other clinically important factors that can only be captured as edge weights.

In a match between pairs at geographically distant transplant centers, the donor of each pair is expected to travel to his actual recipient's transplant center. Not only does this separate family members during the operations and recovery, but donor travel costs are not reimbursed by insurance. Long-distance matches would be restricted to those with the means to fund them, which could cause socioeconomic disparity in access to paired donation. Also, a system that requires most donors to travel would likely discourage participation.

Although kidney donors and recipients need not be related for a good outcome of the transplant, the extent of immunologic concordance between donor and recipient affects survival rates for the kidney. There are six relevant human leukocyte antigens (HLAs) for each person, and reported 5-year survival percentages for a zero-, three-, and six-mismatch live donor kidney are 89%, 74%, and 60%, respectively (Opelz 1997). Analagous differences in survival rates for deceased donor kidneys prompted UNOS to make an exception to its normal allocation rules for these organs. Transplants that would be zero-mismatch are given absolute priority (UNOS 2006c). Still, some members of the transplant community believe that recent advances in immunosuppressive drugs have reduced the importance of closely matched HLA between donor and recipient (Su et al. 2004).

Other factors such as the age and medical history of donor and recipient could be used in addition to HLA matching to generate edge weights that express the expected gain in life-years for each particular donor and recipient pairing. This would, for example, make it more likely that a kidney from a younger donor goes to a younger recipient who can take advantage of the graft's full usable lifespan.

There is an association between sensitivities to HLA antigens in various cross-reactive groups, so that some sensitivities may predict other sensitivities. Using cross-reactive group data, the likelihood of a positive crossmatch between any two people could be predicted prior to a laboratory test. Including this prediction in edge weights would mean that fewer exchanges in the optimal allocation will be scuttled by an unpredicted positive crossmatch. In similar fashion, if a few centers can desensitize recipients to some incompatible donors, edge weights might reasonably depend on the level of difficulty expected in desensitizing recipients to particular donors.

Finally, transplant centers must be assured that feasible matches between pairs who are both at their center will receive priority. Otherwise, large centers with the ability to perform many paired donations among their own patients will have a disincentive to include their patients and donors in a national registry.

In each of the issues above, it is the fit between the donor of the first pair and the recipient of the second pair, and the reverse, that is critical to determining the benefit of an exchange. To incorporate any of the considerations of this section, edge weights must be used to discriminate among matchings that contain different edges but give paired donation opportunities to an identical set of recipients.

## 4. Contrasting Maximum Vertex-Weight Matchings to Maximum Edge-Weight Matchings

In this section, we prove several statements about the number of edges in any maximum vertex-weight or maximum edge-weight matching in a graph. We give our proof of the familiar result that any maximum vertex-weight matching also has maximum cardinality. In contrast, at least one maximum edge-weight system using clinically motivated weights tested by UNOS in a recent computational study resulted in about 3% fewer transplants than were actually possible. (UNOS 2006b). We show that in the worst case, the number of edges in a maximum edge-weight matching may be only half the number of edges in a maximum cardinality matching. Then, we make stronger claims about the number of edges included in a maximum edge-weight matching under the assumption that there are only small differences between any two edge weights. If all the edge weights are close together in a sense we will describe, we prove that a maximum edge-weight matching will have exactly the maximum cardinality. We conclude by suggesting an edge weighting for kidney paired donation that will allow important edge properties to be considered while maximizing the number of transplants performed.

### 4.1. Maximum Vertex-Weight Matchings

The (vertex-) *weighted matching number* $\mu_v$ of $G$ is the maximum number of edges among all matchings with maximum vertex-weight in $G$. The following result assures us that specifying higher priorities to some patients does not decrease the total number of patients receiving transplants.

PROPOSITION 1. *In a vertex-weighted graph with positive vertex weights any matching with maximum vertex-weight also has maximum cardinality. In other words, $\mu_v = \mu$.*

We isolate the main idea of the proof of Proposition 1 in a lemma, which is a reformulation of a fundamental result in graph theory discovered by Berge (Berge 1957, 1972). We let $V(M)$ denote the set of vertices in a matching $M$.

LEMMA 1 (**Matching Lemma**). *Let $G$ be a graph with matching number $\mu$. Let $M^*$ be any matching in $G$. Then there is a matching $M$ with $\mu$ edges such that $V(M^*) \subseteq V(M)$.*

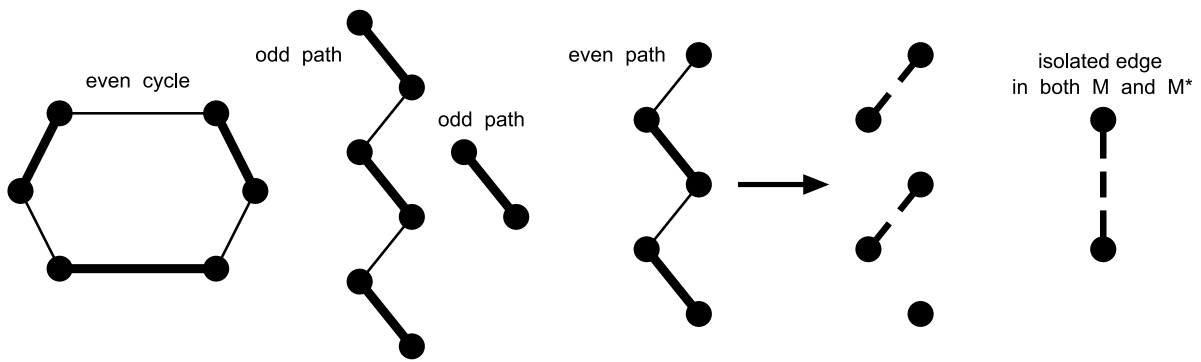

**Figure 2**   Some components of the subgraph H in the proof of the Matching Lemma. The bold edges are in the maximum matching $M$, the unbolded edges are in the matching $M^*$, and the bold dashed edges are in both of the matchings $M$ and $M^*$.

*Proof of the Matching Lemma.*   Let $M$ denote any maximum matching of $G$. We will replace some edges of $M$ by an equal number of edges of $M^*$ to bring about the containment $V(M^*) \subseteq V(M)$. It will not be necessary (and it may not be possible) to include all edges of $M^*$ in a maximum matching $M$. Consider the subgraph $H$ of $G$ whose vertex set is $V(M^*) \cup V(M)$ and whose edge set is $M^* \cup M$. Each connected component of $H$ is either an even cycle or a path with edges alternating between $M$ and $M^*$, or an isolated edge belonging to both $M$ and $M^*$. (See Figure 2.)

The vertices of $V(M^*)$ in any even cycle are already in $V(M)$. In a path with an odd number of edges, the edge at each end must occur in $M$ since $M$ is a maximum matching, and it is clear that $V(M^*) \subseteq V(M)$. In paths with an even number of edges the two matchings use the same number of edges, and we may replace the edges of $M$ by those of $M^*$ (as illustrated in Figure 2) to bring about the containment $V(M^*) \subseteq V(M)$. The vertices of $V(M^*)$ in an isolated edge belonging to both $M^*$ and $M$ are already in $V(M)$. □

The Matching Lemma immediately implies Proposition 1. Let $M_v$ be a maximum vertex-weight matching with $\mu_v$ edges. Then there is a maximum cardinality matching $M$ that satisfies $V(M_v) \subseteq V(M)$. If $V(M_v) \neq V(M)$, then the vertex-weight of $M$ is strictly greater than the vertex-weight of $M_v$, contrary to the definition of $M_v$. Therefore $V(M_v) = V(M)$, and thus $\mu_v = \mu$. The result in Proposition 1 is well-known; see Roth et al. (2005b) for a demonstration in a different context.

### 4.2. Worst-case Cardinality of Maximum Edge-Weight Matchings

The (edge-) *weighted matching number* $\mu_e$ of $G$ is the maximum number of edges among all matchings with maximum edge-weight in $G$. Clearly,

$$\mu_e \leq \mu. \tag{1}$$

Unfortunately, the analogue of Proposition 1 fails. For instance, the edge-weighted graph in Figure 3 has matching number $\mu = 6$ and edge-weighted matching number $\mu_e = 3$. In this example, the ratio $\mu_e/\mu$ equals $1/2$. The following result shows that the ratio cannot be any smaller.

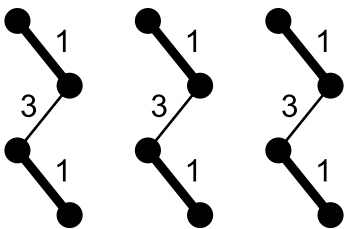

**Figure 3**     A maximum cardinality matching (bold edges) has six edges, while a maximum weight matching has three edges. The labels are the edge weights.

PROPOSITION 2.   *In an edge-weighted graph with positive edge weights, a matching with maximum edge weight has at least half as many edges as a matching of maximum cardinality. In other words,*

$$\mu_e \geq \frac{1}{2}\mu. \tag{2}$$

*Proof.* Let $M_e$ be a maximum edge-weight matching, and let $M$ be a maximum matching of $G$. As in the proof of the Matching Lemma, consider the subgraph $H$ of $G$ with vertex set $V(M_e) \cup V(M)$ and edge set $M_e \cup M$. Again, each connected component of $H$ is either an even cycle or a path. (See Figure 2.) We will show that the inequality $\mu_e/\mu \geq 1/2$ holds for each connected component of $H$. The inequality (2) then follows.

An even cycle satisfies $\mu_e/\mu = 1$, as does an even path. Any odd path in $H$ with $k$ edges has its first, third, ..., $k$th edge in $M$ and its second, fourth, ..., $(k-1)$th edge in $M_e$. There are $(k-1)/2$ edges of $M_e$ and $(k+1)/2$ edges of $M$ on an odd path with $k$ edges, so we have $\mu_e/\mu = (k-1)/(k+1)$. This ratio is at least $1/2$ unless $k=1$. However, an odd path with one edge cannot occur in $H$ since this would correspond to an isolated edge belonging to either $M$ or $M_e$ but not both, violating either the maximum cardinality of $M$ or the maximum edge-weight of $M_e$. $\square$

The prospect of producing a maximum edge-weight allocation with $\mu_e$ near $\mu/2$ is unacceptable. We proceed by restricting the class of edge weightings to obtain stronger guarantees. In Section 4.3 we will assign the edge weights in a manner that guarantees $\mu_e = \mu$.

Note that the preferences specified by vertex weights can always be captured by suitable edge weights. Let $G_v$ be a arbitrary vertex-weighted graph, and let $G_e$ be an edge-weighted graph in which the weight of each edge equals the sum of the weights of the two incident vertices in $G_v$.

PROPOSITION 3. *A matching has maximum vertex weight in $G_v$ if and only if it has maximum edge weight in the corresponding edge-weighted graph $G_e$.*

*Proof.* By the manner in which the edge weights are defined, the vertex weight of any matching in $G_v$ equals the edge weight of the same matching in $G_e$. It follows that a matching has maximum vertex weight in $G_v$ if and only it has maximum edge weight in $G_e$. $\square$

### 4.3. A Theorem on the Cardinality of Maximum Edge-Weight Matchings

If the edge weights of $G$ are all equal, then $\mu_e = \mu$, of course. We will show that the equality $\mu_e = \mu$ still holds provided the edge weights are nearly equal. Moreover, if the edge weights are not nearly equal, there is a lower bound for the fraction $\mu_e/\mu$.

THEOREM 1. *Let $G$ be an edge-weighted graph whose edge weights are all at least $w$ and at most $W$ ($0 < w \leq W$). Let $\mu$ and $\mu_e$ be the matching number and edge-weighted matching number of $G$. Then*

$$\mu_e \geq \left(\frac{w}{W}\right)\mu. \tag{3}$$

*Also, if G has n vertices and $w \neq W$, then*

$$\frac{W}{W-w} > \left\lfloor \frac{n}{2} \right\rfloor \qquad implies \qquad \mu_e = \mu. \qquad (4)$$

*Moreover, when the inequality in (4) holds, then every maximum edge-weight matching has maximum cardinality $\mu$.*

*Proof.* Consider two matchings $M$ and $M_e$ in $G$ with maximum cardinality and maximum weight, respectively. Of course, the weight of $M$ is at most the weight of $M_e$. Also, the weight of $M$ is at least $w\mu$, while the weight of $M_e$ is at most $W\mu_e$. Therefore $W\mu_e \geq w\mu$, and inequality (3) follows.

Because $\mu$ and $\mu_e$ are integers, inequality (3) implies that $\mu_e = \mu$ if $(w/W)\mu > \mu - 1$. This latter inequality is equivalent to $W/(W-w) > \mu$ if we assume that $w \neq W$. In any graph the matching number satisfies $\lfloor n/2 \rfloor \geq \mu$, and thus the implication (4) holds. The argument above remains valid when $\mu_e$ is replaced by $|M_e|$, the cardinality of an arbitrary maximum edge-weight matching. Thus, whenever the premise of (4) holds, every maximum edge-weight matching has maximum cardinality $(|M_e| = \mu)$. $\square$

### 4.4. Edge Weightings Suggested by Theorem 1

Let $G$ be a KPD graph with $n$ vertices, matching number $\mu$, and weighted matching number $\mu_e$. The edge weight restriction of Theorem 1 forces the desirable equality $|M_e| = \mu_e = \mu$. It guarantees that every maximum edge-weight matching reflects a *quantitative efficiency* precept: Any two kidney paired donations are better than any single kidney paired donation.

Theorem 1 points the way to an allocation algorithm for kidney paired donation using the KPD graph $G$. We assign the weights

$$W = n + 1 \text{ to each preferred edge of } G,$$
$$\text{and } w = n - 1 \text{ to each non-preferred edge of } G.$$

An edge is *preferred* provided the two pairs have ready access to the same hospital, say, or the degree of immunologic concordance meets some desired threshold. Because $W/(W-w) = (n+1)/2 > \lfloor n/2 \rfloor$, by (4) we have $\mu_e = \mu$. Then the edges of $M_e$, any maximum edge-weight matching in $G$, give an optimal set of organ exchanges. This allocation maximizes the total number of kidney paired donations while simultaneously reducing travel and increasing immunologic concordance.

In the simplified method presented above, the edge weights take on just two values. No matching with maximum cardinality in $G$ uses more preferred edges than $M_e$. In a more refined model we may specify degrees of preference by assigning edge weights to be real numbers in the closed interval $[n-1, n+1]$. We expand on this idea in Section 5.2.

**4.4.1. A Wider Interval of Edge Weights**   It is not clear whether the medical community will always prefer an allocation that maximizes the total number of transplants. It appears that UNOS may accept a small penalty in overall number of transplants if other allocation objectives are improved (UNOS 2006b). Using (3), we can still offer a guarantee about the size of every maximum edge-weight matching if the clinically appropriate edge weight interval exceeds that in Theorem 1. If ethical considerations allow a matching which is a fraction $1 - \epsilon$ of the maximum cardinality $\mu$, then by inequality (3) we may select our edge weights in any scaled version of the interval $[1 - \epsilon, 1]$.

## 5. Generalizing Beyond The Objectives of Cardinality and Edge-Weight

### 5.1. Maximum Edge-Weight Matchings with Exceptional Recipients

Another reasonable allocation objective is to maximize transplants for a very small group of exceptional recipients. Luckily, this is possible within the constraint of arranging the largest possible number of transplants, and in addition to prioritizing edge properties like immunologic concordance. Surgeons cite the example of patients who have run out of dialysis access and therefore can not be dialyzed. For this group of recipients, a transplant is truly life-saving. In what follows, we consider an edge-weight system that parallels the one in Section 4.3, with the additional specification of an exceptional group of high-priority recipients.

**5.1.1. A Theorem on Edge Weights for Exceptional Recipients.**   Let $G = (V, E)$ be a KPD graph with $n$ vertices. Consider the vertex partition $V = V_1 \cup V_2$, where each vertex in $V_2$ represents an exceptional recipient with his donor(s). We anticipate that $V_2$ will have smaller cardinality than $V_1$. A matching $M$ has *maximum $V_2$-cardinality* provided no matching in $G$ is incident with more vertices of $V_2$ than $M$. Our goal is to assign edge weights to reflect the high priorities of the vertices in $V_2$. Let the edge weight for each edge be the sum of: a number $b > 0$, a number $B > 0$ for each vertex in $V_2$ that is incident with the edge, and a number from the interval $[0, 2]$. Consider the edge partition $E = E_1 \cup E_2 \cup E_3$, where $E_k$ is the set of edges of $G$ with exactly $k - 1$ vertices in $V_2$ for $k = 1, 2, 3$. (See Figure 4.) Then the edge weight of each edge in $E_1$, $E_2$, and $E_3$ lies in the respective closed interval

$$[b, b+2], \quad [b+B, b+B+2], \quad \text{and} \quad [b+2B, b+2B+2]. \tag{5}$$

The number $b$ expresses the priority given to raw cardinality, the number $B$ expresses the priority given to the vertices of $V_2$, and the number in $[0, 2]$ expresses any other desirable properties of a match, say, with larger weights corresponding to shorter travel distance or higher immunologic concordance.

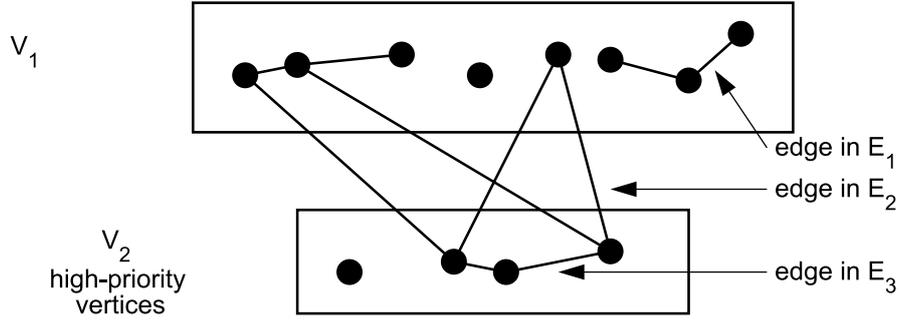

**Figure 4**    Vertex and edge subsets

THEOREM 2. *Let $G$ be an edge-weighted graph with vertex partition $V_1 \cup V_2$, edge partition $E_1 \cup E_2 \cup E_3$, and edge weights in the intervals in* (5), *with $b > 0$ and $B > 0$. Let $G$ have matching number $\mu$ and edge-weighted matching number $\mu_e$.*

(a) *If*

$$b \geq n - 1, \tag{6}$$

*then $\mu_e = \mu$, and every maximum edge-weight matching has maximum cardinality $\mu$.*

(b) *If*

$$B > n, \tag{7}$$

*then every maximum edge-weight matching has maximum $V_2$-cardinality.*

Part (a) asserts that, subject to condition (6) on $b$, every maximum edge-weight matching has maximum cardinality. Part (b) asserts that, subject to condition (7) on $B$, a maximum edge-weight matching also has maximum $V_2$-cardinality. If the premises (6) and (7) are both satisfied, then a maximum edge-weight matching has both maximum cardinality and maximum $V_2$-cardinality. Thus our edge-weighting scheme has desirable properties for kidney paired donation.

*Preliminaries to the proof of Theorem 2.* First consider a vertex-weighted graph $G_v$, where every vertex in $V_1$ has weight $b/2$, and every vertex in $V_2$ has weight $(b/2) + B$. Let $\mu_v$ and $\mu$ be the matching number and vertex-weighted matching number of $G_v$. By Proposition 1, a maximum vertex-weight matching in $G_v$ has maximum cardinality, and so $\mu_v = \mu$. We claim that any maximum vertex-weight matching $M_v$ has maximum $V_2$-cardinality. To see this, note that the Matching Lemma tells us that there is a maximum cardinality matching $M$ with $V(M_v) \subseteq V(M)$. It follows that $M$ has maximum $V_2$-cardinality. Because $B > 0$, $M$ must have greater vertex weight than any maximum cardinality matching that does not have maximum $V_2$-cardinality, and so all maximum vertex-weight matchings must also have maximum $V_2$-cardinality.

We construct $G_e$, the edge-weighted graph corresponding to $G_v$, where the weight of each edge is the sum of the weights of the two incident vertices. By Proposition 3 any maximum edge-weight matching $M_e$ in $G_v$ has maximum vertex weight. From the argument above, $M_e$ also has maximum cardinality and has maximum $V_2$-cardinality. Each edge weight of $G_e$ equals $b$, $b + B$, or $b + 2B$, depending upon the number vertices of $V_2$ incident with the edge.

We now amend the edge weights of the graph $G_e$, adding to each edge any number in the interval $[0, 2]$, to create a new graph $G$. Since the two graphs have the same set of edges, $\mu(G) = \mu(G_v)$. We use the generic $\mu$ to refer to the identical matching number of $G$ and $G_v$. We will show that $\mu_e(G) = \mu$ if (6) holds, and that any maximum edge-weight matching in $G$ has maximum $V_2$-cardinality if (7) holds.

*Proof of* (a). We proceed by contradiction. Assume that $\mu_e(G) < \mu$. Then because $\mu_e$ and $\mu$ are integers, $\mu_e(G) \leq \mu - 1$. Let the weight of every maximum edge-weight matching in $G_e$ be $K$. The weight of a maximum edge-weight matching in $G$ is at most $K - b + 2(\mu - 1)$, because it has one fewer edge (subtracting at least $b$) but may gain as many as $2(\mu - 1)$ units of weight from the amended weights of $G$. But by assumption (6), $K - b + 2(\mu - 1) \leq K - n - 1 + 2\mu \leq K - n - 1 + 2\lfloor n/2 \rfloor < K$. The last inequality holds because $2\lfloor n/2 \rfloor \leq n$. Then the weight of a maximum edge-weight matching in $G$ is strictly less than the weight of every maximum edge-weight matching in $G_e$, even though no edge weight in $G_e$ exceeds the corresponding edge weight in $G$. This is a contradiction.

*Proof of* (b). We proceed by contradiction. Assume that some maximum edge-weight matching $M_e(G)$ does not have maximum $V_2$-cardinality. Again let the weight of every maximum edge-weight matching in $G_e$ be $K$. The weight of a maximum edge-weight matching in $G$ is at most $K - B + 2\mu$, because it contains at least one fewer vertex from $V_2$ (subtracting at least $B$), and may gain as many as $2\mu$ units of weight from the amended weights in $G$. But by assumption (7), $K - B + 2\mu < K - n + 2\mu \leq K - n + 2\lfloor n/2 \rfloor \leq K$. Again the weight of a maximum edge-weight matching in $G$ is strictly less than the weight of every maximum edge-weight matching in $G_e$, and we have a contradiction. $\square$

## 5.2. Edge Weightings for Clinical Paired Donation Registries.

Deciding what constitutes the best allocation is outside the scope of mathematics. Those determinations must be made by the transplant community in a manner that reflects both judgment and values, as discussed in Section 2.4. UNOS has already narrowed the list of potential considerations in weighting the edges to the following: geography (same transplant center, same state, or same

UNOS region), zero-mismatches, prior live donor recipients, pediatric recipients, sensitized recipients, and waiting time. Policy-makers at UNOS have hand-tuned various edge-weight schemes using simulated KPD graphs to arrive at their current proposal (UNOS 2006b). Theorem 1 and Theorem 2 provide some guarantees regarding the effect of selecting particular numerical weights. If the correct clinical model is to order the cardinality objective(s) strictly ahead of other objectives in the ways we have described, then policy-makers may safely select the edge weights from the specified intervals. Our theorems guarantee that the maximum edge-weight matchings solve the corresponding preemptive multi-objective problems.

The results presented here suggest a simple calculation to assign edge weight $w_i$ to edge $i$ in practice. Initially, assign a weight $\tilde{w}_i$ in the closed interval $[0, 2]$ to edge $i$. The number $\tilde{w}_i$ may be assigned on the basis of edge properties, such as the relative medical and geographic desirability of particular matches. Also, $\tilde{w}_i$ may be assigned on the basis of vertex properties, such as whether recipients are pediatric or prior live donors, when these recipients are not classified as exceptions.

Once $n$ (the total number of patient-donor pairs) is known, add $b = n - 1$ to each edge weight $\tilde{w}_i$ to get the weight $w_i$ for edge $i$. If the consensus among physicians is that quantitative efficiency is required, then each edge weight falls in the interval $[n - 1, n + 1]$. The advantageous conclusion of Theorem 1 applies.

On the other hand, say that the desired allocation does not require quantitative efficiency, but a subset $V_2$ of exceptional vertices has been identified. Then we add $B = n + 1$ to each edge weight $\tilde{w}_i$ for each vertex in $V_2$ incident with edge $i$. Thus either $0$, $n + 1$, or $2n + 2$ is added to $\tilde{w}_i$. Because no edge should have weight zero, we add a small number to each edge, say, 1. The resulting edge weights are within the intervals

$$[1, 3], \qquad [n + 2, \, n + 4], \qquad \text{and} \qquad [2n + 3, \, 2n + 5]. \tag{8}$$

These edge weights might be suitable if there are some exceptional recipients, but policy-makers are willing to trade off overall cardinality to achieve, say, greater reductions in travel. Then part (b) of Theorem 2 applies, and as many exceptional recipients as possible will be matched.

If physicians require both quantitative efficiency and the maximum number of transplants for exceptional recipients, then we add $n - 1$ to every edge, and we also add $B = n + 1$ to each edge weight $w_i$ for each vertex in $V_2$ incident with edge $i$. The edge weights will fall in the intervals

$$[n - 1, \, n + 1], \qquad [2n, \, 2n + 2], \qquad \text{and} \qquad [3n + 1, \, 3n + 3], \tag{9}$$

and both of the reassuring conclusions of Theorem 2 hold. Donors and recipients can have confidence in an organ allocation system that maximizes the number of people who receive a transplant, and that gives absolute priority to exceptionally deserving recipients, and that also takes into account other important concerns.

### 5.3. Restricting Edge Weights Is a Multi-Objective Optimization Method

It is instructive to recast Theorem 2 and the discussion in Section 5.2 as a preemptive (lexico-graphical) multi-objective optimization method. Klingman and Phillips (1984) construct a similar method for a related model, a preemptive multi-objective assignment model.

Recall that the number of vertices in a matching $M$ is $|V(M)|$. Let $|V_2(M)|$ be the number of vertices of $M$ that are in the subset $V_2$. In decreasing order of importance, the objectives we consider for kidney paired donations are: $f_1(M) = |V_2(M)|$, the number of transplants for the highest priority recipients; $f_2(M) = |V(M)|$, the number of transplants overall; and $f_3(M)$, a function that may subsume various other vertex-associated objectives, such as the number of transplants for preferred but not exceptional recipients, as well as edge-associated objectives, like travel and immunologic concordance.

Let $\mathbb{M}$ be the set of all matchings in $G$. We express the constraints in the generic form: $M \in \mathbb{M}$. The general theory of multi-objective combinatorial optimization asserts (Ehrgott and Gandibleux 2000) that for an ordered list of objectives $f_1(M)$, $f_2(M)$, …, $f_k(M)$ there exist preemptive weights $C_1$, $C_2$, …, $C_k$ such that the preemptive multi-objective problem is equivalent to

$$\max_{M \in \mathbb{M}} \sum_{i=1}^{k} C_i f_i(M). \tag{10}$$

One may solve (10) using a maximum edge-weight matching algorithm.

Without loss of generality $C_k = 1$, and in our case $k = 3$. Theorem 2 provides exact values for $C_1$ and $C_2$. Write the maximum edge-weight matching problem as

$$\max_{M \in \mathbb{M}} \sum_{j \in M} w_j = \max_{M \in \mathbb{M}} \left( (n+1)|V_2(M)| + (n-1)|V(M)| + \sum_{j \in M} \tilde{w}_j \right) \tag{11}$$

and note that the objective on the right hand side of (11) is a linear combination of the objectives $f_1$, $f_2$, and $f_3$. This yields $C_1 = n+1$ and $C_2 = n-1$ as preemptive weights for (10).

An alternative method for solving preemptive multi-objective problems is as a sequence of single-objective problems. In that paradigm, first solve

$$F_1 = \max_{M \in \mathbb{M}} f_1(M) \tag{12}$$

and then solve

$$F_2 = \max_{M \in \mathbb{M}} f_2(M), \text{ subject to } F_1 = f_1(M), \tag{13}$$

which incorporates maximizing the most important objective as a constraint, and then solve,

$$F_2 = \max_{M \in \mathbb{M}} f_3(M), \text{ subject to } F_1 = f_1(M) \text{ and } F_2 = f_2(M), \tag{14}$$

and so on. The advantage of the former method which combines the objectives into a single one is that an efficient algorithm for matching can be used, whereas there is no obvious way to incorporate side constraints of the form in (13) into a maximum matching algorithm.

The structure of many organ allocation methods is hierarchical: for instance, if multiple zero-antigen mismatches are found, a deceased donor kidney goes first to a zero-mismatch local candidate, second to a zero-mismatch sensitized candidate in an area that is owed a payback from an earlier zero-mismatch transfer, third to a zero-mismatch sensitized candidate within the larger region, and so on. A hierarchical policy for kidney paired donation could be implemented using a hierarchical collection of subsets, say, $V_1, V_2, \ldots, V_k$, and an appropriate edge-weighting. However, we do not know the form of the $C_i$ in general. Because preemptive weights that convert a multi-objective maximum edge-weight matching problem into a single-objective one are known to exist, an interesting and practical generalization of our results would be to find the $C_i$.

## 6. Computational Trials

### 6.1. Solution Algorithms

There are polynomial-time algorithms (e.g., Galil et al. (1986), Lovász and Plummer (2009)) to produce a matching with maximum cardinality, or vertex weight, or edge weight, in an undirected graph. In each case, the optimal matching need not be unique. The enumeration of all matchings with maximum cardinality or weight is a computationally difficult problem (Valiant 1979). Thus examining all maximum cardinality matchings is not an attractive strategy for satisfying a secondary objective subject to maximum cardinality. The size of realistic kidney paired donation problems is fairly modest. Based on medically detailed simulations, we predict that in the most optimistic scenarios about 750 patient-donor pairs would arrive each quarter to a national KPD program (Gentry et al. 2005). A maximum edge-weight matching for a 750-vertex KPD graph can be found in a few seconds using, say, the commercial graph package LEDA from Algorithmic Solutions Software GMBH (Mehlhorn et al. 1997).

In contrast, a straightforward encoding of optimal two- and three-way allocation using CPLEX optimization software to solve the integer program (15) detailed in Section 7.3 fails because there are on the order of a million two- and three-way cycles in a 750-vertex directed KPD graph. We are aware of recent progress in specialized heuristics and column generation techniques to solve these large-scale integer programs (Abraham et al. 2007).

### 6.2. Computational Experiments

We used clinically realistic simulations of incompatible pairs to test our suggested edge-weight methods against a vertex-weight method (*Vertices*) and the proposed UNOS edge weights (*UNOS* *proposal*). We compared the optimal allocations determined by these weighting methods, using 100 pools of 750 incompatible donor-recipient pairs. Results are shown in Table 1. Each row shows outcomes for a distinct weighting scheme, and the columns in Table 1 reflect the average numbers of: incompatible pairs that can be matched for transplantation; incompatible pairs who match to another pair in their UNOS region and thus do not need to travel; zero-mismatch transplants; prior live donor transplants; pediatric transplants; and sensitized transplants. The average number of pairs of each type who were matched for five paired donation allocation schemes should be compared to the numbers in the **Totals** row, which give the average number of pairs of each type in the KPD graphs generated. As an example, the average number of pediatric recipients was about 22 out of the 750 recipients. It would not be meaningful to provide the number of edges having certain edge properties in this row.

The columns of the table correspond to edge properties (local and zero-mismatch) and vertex properties (prior donor, pediatric, and sensitized) that are relevant to allocation for kidney paired donation, along with the total number of people transplanted. Each of the columns is a separate outcome measure, and larger values are more desirable in every column. Not shown are the numerical weights assigned to the properties in the columns, which differed between rows. The numerical weights assigned for each property that pertained to a particular edge, or a vertex incident with that edge, were summed to get the weights for each edge. The weights for edge properties were zero in the *Vertices* scheme.

The *Edges* 1 method was suggested in Section 4.4, and it yields maximum cardinality allocations that favor local and zero-mismatch edges. The *Edges* 2 method is to weight edges in the intervals (8) of Section 5.2, and it yields a maximum $V_2$-cardinality matching that does not necessarily have maximum cardinality. The *Edges* 3 method is to weight edges in the intervals (9) of Section 5.2, and it yields maximum cardinality and maximum $V_2$-cardinality allocations that favor local and zero-mismatch edges. The subset $V_2$ is composed of: the very small group of prior live donors (about 0.05% of the total); the group of pediatric recipients (about 3%); and the large group of sensitized recipients (about 28%). The latter group of sensitized recipients differs from the average in that sensitized recipient vertices have very few incident edges. Pediatric and prior live donor candidates are without connectivity bias.

The results suggest that vertex-weighting schemes are impractical because they require too many people to travel long distances. The vertex-weighting shown is weakly Pareto-dominated by our edge-weighting schemes. The *Edges* 1 weighting achieves the greatest number of local transplants but, since it gives the vertex properties zero weight, it matches fewer pediatric and other exceptional recipients. The UNOS proposal does not achieve the maximum number of transplants, but it does not miss the optimum by very much, and it does achieve more zero-mismatch transplants than any other weighting scheme shown. This is typical of multi-objective problems; relaxing a requirement to find the absolute maximum of one objective (cardinality or $V_2$-cardinality) allows a rise in other objective values. The *Edges* 2 weighting is similar to the UNOS weighting in that it trades off maximum cardinality for other objectives, but *Edges* 2 gives up more total transplants in order to achieve the maximum number of transplants for the exceptional recipients. We remark that our research group contributed substantially to the UNOS proposal for edge weightings, so the proximity between the *UNOS proposal* row of the table and our rows is not coincidental.

The maximum edge-weight matching for any scheme in which all properties have positive weight is on the efficient frontier for this allocation problem (Ehrgott and Gandibleux 2000). Starting from any of the schemes shown in which all listed properties have positive weight (these are *Edges* 2, *Edges* 3, and *UNOS*), it will not be possible to increase one outcome measure without decreasing some other outcome measure.

## 7. Generalizations of Kidney Paired Donation

We introduce some variants of kidney paired donation. Many generalizations of the classic kidney paired donation between two incompatible pairs have been proposed, mostly in response to the difficulty of finding complementary pairs in the small registries that now exist. Some of these alternative donation arrangements require only trivial alteration of the underlying KPD graphs, but in some cases allocation is no longer equivalent to matching.

### 7.1. Compatible Paired Donation

Voluntary compatible paired donation (Gentry et al. 2007, Roth et al. 2005a) allows recipients who have compatible donors to enter paired donation arrangements with incompatible pairs. The match rate for incompatible pairs would nearly double if compatible pairs participated in paired donation, because the compatibles could offset the blood type imbalance among incompatible pairs (Gentry et al. 2007). Compatible paired donation has also been referred to as altruistically unbalanced exchange (Ross and Woodle 2000), but we would argue that it need not be altruistically unbalanced. Recipients with compatible donors might participate in a paired donation registry in order to find a younger or more closely-matched donor, while others may indeed participate out of altruistic motivation.

A KPD graph may include compatible pairs among its vertices without altering any of the development in this paper. Compatible pairs could proceed with direct donation to the intended recipient, which suggests that a reflexive edge (self-loop) be attached to each compatible pair vertex. Equivalently, each compatible pair vertex could be attached to its own dummy vertex representing direct donation.

## 7.2. List Paired Donation

A New England consortium has arranged a number of list paired donations, in which a recipient's incompatible donor gives to the person at the top of the deceased donor waiting list, and the intended recipient is moved to the top of the waiting list for the next compatible deceased donor organ (Delmonico et al. 2004). List paired donation can be represented in a KPD graph by connecting a self-loop, or a dummy edge and vertex, to each incompatible pair who is eligible for list paired donation.

Ethical objections to list paired donation have been raised. Blood type O donors are scarce among incompatible pairs, so list paired donation will most often reallocate a type O donor away from the waiting list in exchange for a donation from a non-O live donor to the waiting list. This non-O for O list donation seems unfair to people waiting for a type O deceased donor, who would see a disproportionate increase in waiting times (Zenios et al. 2001). If ethical restrictions prevent non-O for O list paired donation, and practical restrictions prevent list paired donation for sensitized recipients, then nearly everyone who could participate in list paired donation could instead be matched in a national live paired donation registry (Gentry et al. 2005). Live paired donation is preferable to list paired, because kidney allografts from living donors are expected to function almost twice as long (median survival of 21 years versus 13 for deceased donor organs Harihan et al. (2000)). We expect that on a national level, live paired donation will supersede list paired donation.

## 7.3. Donations Among Three or More Pairs

To model allocations that allow participation of more than two pairs in a donation we use a *directed* KPD graph $G$. There is an edge of $G$ from one vertex to another vertex provided the donor of the source vertex is compatible with the recipient of the target vertex. Then any directed cycle in $G$ is an opportunity for an exchange. We continue to refer to an $k$-way exchange of kidneys as paired donation, because the incompatible donor and his intended recipient are *paired*.

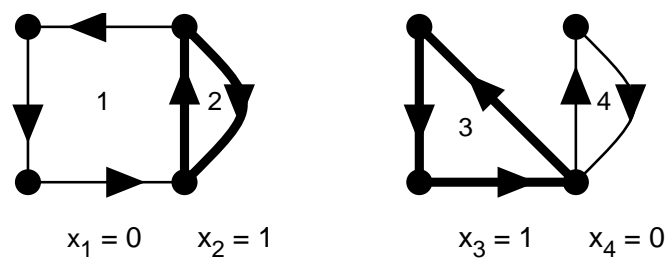

**Figure 5**    A directed KPD graph with four directed cycles, two of which are selected in an allocation (bold edges).

When more than two pairs are allowed in an exchange, the allocations are not matchings, but rather sets of vertex-disjoint directed cycles in the directed KPD graph $G$. An optimal allocation can be found using integer (binary) programming. A simple integer program for optimal allocation, allowing participation of at most $k$ pairs in each exchange, proceeds as follows. Suppose $G$ has $n$ vertices, and $p$ directed cycles with at most $k$ edges. With the $i$-th directed cycle we associate the binary decision variable $x_i$. Let $x_i = 1$ if the $i$-th cycle is included in the allocation, and let $x_i = 0$ if the $i$-th cycle is not included in the allocation, as in Figure 5. Let the cycle index $i$ be in the set $\text{Cyc}(j)$ provided the $i$-th cycle contains vertex $j$ $(j = 1, 2, \ldots, n)$. Let the weight $w_i$ represent the value of including the $i$-th cycle in the allocation. Then the integer program to be solved is

$$
\begin{aligned}
\max & \sum_i^p w_i x_i \\
\text{subject to:} & \sum_{\text{Cyc(j)}} x_i \le 1 \quad \text{for each vertex } j \text{ in } \{1, 2, \ldots, n\}, \\
& x_i \in \{0, 1\} \quad \text{for each cycle } i \text{ in } \{1, 2, \ldots, p\}.
\end{aligned}
\tag{15}
$$

Exchanges involving three donor-recipient pairs have been accomplished at a few transplant centers (McLellan 2003, Montgomery et al. 2005). Saidman et al. (2006) have shown that allowing two- and three-way matches would increase the number of transplants that can be performed, especially in small pools of incompatible pairs. Under the assumption of large KPD graphs where blood type incompatibilities but no tissue incompatibilities exist, Roth et al. (2007) have produced some nice results. First, it is possible to achieve the maximum number of transplants without any trades larger than four pairs. Second, Roth et al. (2007) have calculated an upper bound, based on the prevalence of different blood type combinations, on how many transplants are foregone in restricting to two- and three-way trades and how many transplants are foregone in restricting to two-way trades.

For unrestricted $k$-way transfers of kidneys, Roth et al. (2004) suggest that allocation could be accomplished via a modified top trading cycles mechanism, in which every recipient ranks the available kidneys from most to least desirable. However, unrestricted $k$-way matches have been rejected by the medical community, both because of the logistical complexity of simultaneous transplants and the high likelihood of scuttling entire exchanges because of a medical event that affects one donor or recipient in the cycle.

### 7.4. Domino Paired Donation

Altruistic individuals sometimes offer to donate a kidney to any recipient who could benefit, and these individuals are called non-directed donors. Non-directed kidney donations have often been allocated to the recipient on the waiting list for a deceased donor kidney. Instead, in a domino paired donation (Montgomery et al. 2006), also called a chain (Roth et al. 2006), the non-directed donor gives to a recipient with an incompatible donor, and that donor in turn gives to a recipient on the waiting list. The recipient with the incompatible donor benefits because he would not receive the non-directed donation otherwise, and overall the non-directed donor makes two transplants possible. Domino paired donations can also be extended to chains of three or more recipients, and surgeons at Johns Hopkins Hospital recently performed a five-way domino operation. A non-directed donor can be represented as a vertex in either a directed or an undirected KPD graph, where the non-directed donor's "recipient" will be a recipient on the waiting list who is compatible with the donor at the end of the domino.

## 8. Discussion and Limitations

We will review some limitations of our analysis. An important limitation is that our theoretical results about the size of maximum edge weight matchings on undirected graphs apply only to two-way paired donation systems. It is an open question whether there exist edge weights for $k$-way exchange graphs for which maximum edge-weight allocations are also certain to have maximum cardinality. That guarantee could be made in a different way for $k$-way or even two-way matchings, by first calculating the maximum cardinality of a $k$-way or two-way matching, and then using that number as a lower bound on the number of transplants in (15) as a side constraint. Although this would give an allocation with the required property, one could no longer interpret the result as giving an edge weight system that correctly reflects the allocation objective.

No consensus on the allocation objective has been reached in the transplant community, not even to specify that maximizing the number of transplants is required. Still, an allocation method that maximizes the number of transplants is likely to be more palatable to donors and recipients than one that does not.

Polynomial-time algorithms for maximum edge weight matching are well-known in the computer science and optimization literature. However, understanding matching algorithms requires some familiarity with optimization and primal-dual formulations. This unfortunately means that the workings of any optimal allocation procedure will not be accessible to most transplant professionals, which might generate some resistance to these methods. Even ascertaining the edge weights to express a desired allocation objective will be a thorny problem, although the results we have presented constitute a partial answer. In contrast, the allocation of deceased donor organs is much more transparent. Because organs harvested from a deceased donor cannot be stored, each organ is allocated in greedy fashion to the recipient with the largest number of points.

If a KPD graph has smallest edge weight $w$ and largest edge weight $W$, inequality (3) guarantees that $\mu_e \geq (w/W)\mu$. For realistic paired donation graphs, this lower bound is extremely conservative. For instance, in the simulation results of Table 1 for *UNOS* and *Edges* 2 weightings, one cannot use inequality (3) to give tighter bounds than (2) gives, because for the aforementioned weightings the fraction $w/W$ is less than $1/2$. These matchings nonetheless have almost maximum cardinality because they satisfy $\mu_e \geq 0.99\mu$.

Laboratory crossmatch testing of all recipients to their potentially compatible donors in a paired donation registry would be too expensive and difficult to undertake. Even using state-of-the-art techniques to predict all of the sensitivities of each recipient, unpredicted positive crossmatches will still be discovered after the maximization step. This means that any fixed allocation, including the computed optimal allocation, may not be realizable. After laboratory screening of the exchanges in the initial allocation, any pairs in scuttled exchanges with positive crossmatch tests can be restored to the registry along with pairs that were unmatched initially, for a second and possibly a third or fourth allocation step.

We have not discussed the incentive structure for individual recipients and their incompatible donors within a kidney paired donation allocation system. A strategy-proof allocation mechanism is one in which no participant can gain an advantage by misrepresentation. Under the assumption that participants are indifferent among all acceptable kidneys, maximum edge-weight matching is a strategy-proof mechanism. This result follows the argument in Roth et al. (2005b), as generalized by Hatfield (2005).

It is critical that optimization researchers develop and disseminate appropriate allocation methods for kidney paired donation. Transplant communities in the U.S. and other countries are making rapid progress toward national kidney paired donation systems. Allocation methods that may find suboptimal or non-efficient allocations are known to have been used in practice. Also, physicians may not be expert at translating their judgments about the relative value of particular paired donation matches to numerical weights. Our broader aim is to help the transplant community express its allocation goals through the use of KPD graphs. To that end, we have provided weighting methods that give cardinality guarantees, while allowing the necessary edge preference structure. These weighting methods can be seen as specifying the preemptive weights for converting a preemptive multi-objective optimization model to a single-objective model. Following on the work presented here, we are seeking simpler means for eliciting physicians' value judgments about allocation. We hope to supplement the trial-and-error simulations that have permeated clinical dialogue (UNOS 2006b) on the design of kidney paired donation allocation systems.

## Acknowledgments

Sommer Gentry was supported by the the Naval Academy Research Council. Dorry Segev was supported by American Society for Transplantation Clinical Science Faculty Development Award. Both thank Robert Montgomery, M.D., D.Phil. for his leadership and vision in advancing kidney paired donation. Sommer Gentry thanks Fuhito Kojima for a helpful conversation about incentives in paired donation and Charles Mylander for a thoughtful review of the manuscript. We have been pleased to work with Ken Andreoni, M.D., and the United Network for Organ Sharing's kidney-pancreas committee on formulating clinically relevant optimization models for kidney paired donation. We acknowledge the editors and reviewers for their valuable suggestions.

**Table 1**    Comparison of weighting methods for kidney paired donation allocation, using simulated donor-recipient pairs.

| Weighting | Transplants | Local | Zero-mismatch | Prior Donor | Pediatric | Sensitized |
|---|---|---|---|---|---|---|
| **Totals** | 750 | | | 0.37 | 22.36 | 211.64 |
| Edges 1 | 341.58 | 273.14 | 3.80 | 0.15 | 10.19 | 22.33 |
| Edges 2 | 339.42 | 263.42 | 3.97 | 0.27 | 16.61 | 27.91 |
| Edges 3 | 341.58 | 261.38 | 3.69 | 0.27 | 16.61 | 27.91 |
| Vertices | 341.58 | 40.64 | 0.71 | 0.27 | 16.61 | 27.91 |
| UNOS proposal | 341.06 | 263.94 | 4.06 | 0.26 | 16.10 | 25.60 |